\documentclass[referee]{aa} 

\usepackage{subfigure}
\usepackage{graphicx}
\usepackage{txfonts}
\usepackage{xcolor}
\usepackage{threeparttable}
\begin{document}

   \title{Applying saliency-map analysis in searches for pulsars and fast radio bursts}

   \author{C. Zhang
          \inst{1,2,3,4}
          \and
          C. Wang\inst{3}
          \and 
          G. Hobbs\inst{4}
          \and
          C. J. Russell\inst{5}
          \and
          D. Li\inst{2,6,7}
          \and
          S.-B. Zhang\inst{2,4,8}
          \and
          S. Dai\inst{4}
          \and
          J.-W. Wu\inst{1}
          \and
          Z.-C. Pan\inst{1}
          \and
          W.-W. Zhu\inst{1}
          \and
          L. Toomey\inst{4}
          \and
          Z.-Y. Ren\inst{1}
          }

   \institute{National Astronomical Observatories, Chinese Academy of Sciences, A20 Datun Road, Chaoyang District, Beijing 100101, China\\
              \email{zhangchao215@mails.ucas.ac.cn}
         \and
             University of Chinese Academy of Sciences, Beijing 100049, China 
             \and
             CSIRO Data61, Sydney, NSW 2015, Australia
             \and
             CSIRO Astronomy and Space Science, Australia Telescope National Facility, Box 76, Epping, NSW 1710, Australia
             \and
             CSIRO Scientific Computing, Sydney, NSW 2015, Australia
             \and
             CAS Key Laboratory of FAST, National Astronomical Observatories, Chinese Academy of Sciences, Beijing, P.R. China, 100012
             \and
             NAOC-UKZN Computational Astrophysics Centre (NUCAC), University of KwaZulu-Natal, Durban 4000, South Africa
             \and
             Purple Mountain Observatory, Chinese Academy of Sciences, Nanjing 210008, China
             }

   \date{Received September 15, 1996; accepted March 16, 1997}


  \abstract
   {To investigate the use of saliency-map analysis to aid in searches for transient signals, such as fast radio bursts and individual pulses from radio pulsars.}
   {We aim to demonstrate that saliency maps provide the means to understand predictions from machine learning algorithms and can be implemented in piplines used to search for transient events.}
   {We have implemented a new deep learning methodology to predict whether or not any segment of the data contains a transient event.  The algorithm has been trained using real and simulated data sets. We demonstrate that the algorithm is able to identify such events. The output results are visually analysed via the use of saliency maps.}
   {We find that saliency maps can produce an enhanced image of any transient feature without the need for de-dispersion or removal of radio frequency interference. Such maps can be used to understand which features in the image were used in making the machine learning decision and to visualise the transient event. Even though the algorithm reported here was developed to demonstrate saliency-map analysis, we have detected, in archival data, a single burst event with dispersion measure of $41$\,cm$^{-3}$pc that is not associated with any currently known pulsar.}
   {}


   \maketitle
%

\section{Introduction}

Radio telescope observing systems continue to be used to record high-time resolution data sets.  In such data sets the total intensity of the received radio signal is sampled typically every $\sim 100$\,$\mu$s and with moderate (e.g., $\sim$\,MHz) channel bandwidths.  For most historical data sets the samples are 1 or 2 bit digitised, but for many current surveys higher-bit data streams are recorded. High time resolution data sets are used to search for pulsars by seeking for weak periodic signals within the data. They are also used to search for bright, individual pulses from pulsars and fast radio bursts (FRBs) \citep{2016MNRAS.459.1519D,2018MNRAS.480.3457M,10.1093/mnras/stz1748,2012MNRAS.422..379B,2018AJ....156..256C}.

\begin{figure}
    \centering
	\includegraphics[width=1.0\columnwidth,viewport=0 0 425 320, clip]{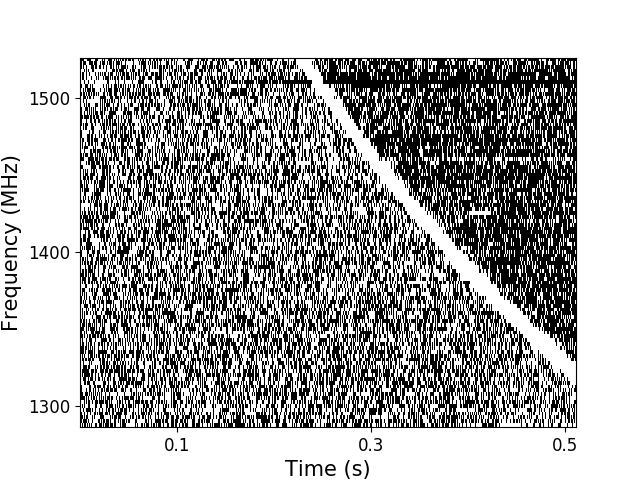}
    \caption{An example of part of the data file containing the original FRB event, the ``Lorimer burst''. Each sample is one bit sampled, with white positive and black negative.}
    \label{fig:imageFRB}
\end{figure}

FRBs are bright, millisecond-duration radio transients. The observed pulses are characterized by dispersion measures (DMs) that are significantly larger than the expected Milky Way contribution. They have been detected at flux densities between tens of micro-janskys and tens of janskys
\citep{2007Sci...318..777L,2016Natur.531..202S,2018AJ....156..256C,2019Natur.566..235C}.  Understanding the origin of FRBs is still an active research area, with many different theoretical explanations \citep{2018arXiv181005836P}. The first FRB was discovered by \cite{2007Sci...318..777L} during the reprocessing of archival pulsar survey data, and it is now commonly referred to as the ``Lorimer burst''. A small segment of the data stream that was used to discover the ``Lorimer burst'' is shown in Figure~\ref{fig:imageFRB}. This particular data file has 96 frequency channels (shown on the y-axis) spanning a total bandwidth of 288\,MHz and each time/frequency sample (also referred to here as a ``pixel'') has been 1\,bit sampled.

Most of the known FRBs have only been detected once.  However, there is now a small population of FRBs in which repeating signals have been detected  (\citealt{2016Natur.531..202S}, \citealt{2019Natur.566..235C} and \citealt{2019arXiv190803507T}).   There are likely thousands of detectable events each day across the full sky, but only a relatively small number have been published to-date. This is because of the moderate field-of-view that many radio telescopes have.  Wide field-of-view telescopes such as The Canadian Hydrogen Intensity Mapping Experiment and the Australian Square Kilometre Array Pathfinder are now operating  and a large number of FRB events will soon be published.   However, for each detected FRB event, current survey processing methods usually produce thousands of false-positive triggers \citep{2018AJ....156..256C}.  Some of these candidates can be rejected based on extra information, such as detection in multiple observing beams, but many of the diagnostic plots are simply inspected visually \citep{2018AJ....156..256C}.  Future telescopes, such as the Square Kilometre Array, will carry out pulsar and FRB searches, but the enormous data rate from those telescopes implies that real-time processing is likely to be required.  Real-time processing methods already operate for FRB searches (e.g., \citealt{2012MNRAS.422..379B}), but produce large numbers of candidates (most of which are false-positive candidates).

Machine learning algorithms are increasingly taking a role in deciding which signals to record for further analysis. Ways to minimise their false positive rates, to maximise their efficiency and their robustness in the presence of radio frequency interference (RFI) need to be explored. \cite{2018MNRAS.480.3457M} designed a machine-learning classifier to identify single-pulses in a strong RFI environment, which relies on features such as the pulse width, DM, and signal-to-noise (S/N) and has been used to discover seven pulsars. \cite{10.1093/mnras/stz1748} detected five new FRBs in real-time with the Molonglo Radio Telescope. The pipeline adds an additional stage to the HEIMDALL pipeline \citep{2012MNRAS.422..379B} to classify the resulting candidates using features extracted from frequency-time data.  \cite{2018AJ....156..256C} focused on reducing the false positive rate from candidates obtained using a traditional search method. They applied a deep learning method to single pulse classification and developed a hierarchical framework for ranking events by their probability of being true astrophysical transients. \cite{2018ApJ...866..149Z} presented the first successful application of deep learning to direct detection of fast radio transient signals in raw frequency-time data. They found 72 new pulses from the repeating fast radio burst FRB 121102 using the Green Bank Telescope.

All these algorithms report on whether a particular candidate (or image segment) contains a likely astrophysical burst event, or not.  However, they generally \emph{do not provide information on what parameters, or what features in the image were used to make that decision}, and so deep learning models are often criticized as ``black boxes'' for decision making. This is primarily because the results from non-linear fitting of the high dimensional data is difficult to  explain intuitively. A \emph{pulse verification process} is often needed to identify real signals from candidates produced by the machine learning modules. Many of these processes use visual inspection and assume that signals can be visually detectable directly by humans~\citep{2018ApJ...866..149Z}. In the presence of RFI or complex noise patterns, direct visual inspection is likely to miss real signals. We address this problem by ensuring that the machine learning procedure not only predicts whether an event is present, or not, but also provides information on how it came to that decision.   This line of work is catagorized as machine learning interpretability. 

There have been many efforts attempting to address the problem of understanding why a machine learning method has made a decision and and to provide users with more confidence in the model predictions (see \citealt{MONTAVON20181}).  For our purpose, we wish to identify the part of an input image that a classifier has identified as being from an astrophysical burst event and we make use of ``saliency analysis'' for this purpose. 

Saliency map analysis has been used for numerous ML applications to highlight features in an input that are relevant to the predictions of a model ~\citep{simonyan2013deep,Sundararajan:2017:AAD:3305890.3306024}, but, to the best of our knowledge, has not been used to date on high-time resolution radio astronomy data sets to render the features of transient events.  

The work described here is explicitly linked to the saliency-map analysis and can be applied to the output of any deep learning method used to search for transient events. Of course, to demonstrate saliency-analysis we require a deep learning method and test data sets, but note that saliency analysis is generally applicable and we do not, here, attempt to convince the reader that our method is better or worse than any other existing algorithms for the FRB searching. In \S2 we describe our machine learning algorithm, training procedure and our test data set that we have used for this work.  We describe the saliency analysis method in \S3 and conclude in \S4.  

\section{FRB classifier and data set}
\subsection{Data Set}
To demonstrate the saliency analysis we make use of archival data sets that are publically available from the Parkes data archive (\url{ https://data.csiro.au}; see \citealt{2011PASA...28..202H}). The data sets are used both to train the algorithm and to provide example observations (containing known pulsars, FRBs and radio frequency interference) to test the effectiveness of the procedure.

All data sets were obtained with the Parkes telescope using the 21-cm multibeam receiver. The primary goal for carrying out the original observations was to search for new pulsars \citep{2006ApJ...649..235M}.  The data files are in PSRFITS \citep{2004PASA...21..302H} search mode format. They are two dimensional spectrograms (time versus frequency) that span a frequency range from 1231 to 1516\,MHz with 96 frequency channels. The time sampling varies between the different data files (from 125\,$\mu$s to 1\,ms). The archive contains more than 100 observing projects, with each observing semester for each project stored as a data collection. Approximately 600 such data collections are now available for public access. Even though the observations were processed by the original science teams, new discoveries are still being made based on these archival data sets (recent discoveries have been reported by Pan et al. 2016, Zhang et al. 2018a and \citealt{2019MNRAS.484L.147Z}).

A single observation often contains millions of time samples. We cannot simply pass the entire data file as an image into a machine learning classifier as (1) typical algorithms require much smaller image sizes and (2) the signals of interest (that is, the astronomical bursts) only last for very short time durations and hence make up a tiny fraction of the entire observation.  We therefore take each observation and split the file into small segments (for this work we choose segments of 512 time samples).  We classify each segment into two categories:

\begin{enumerate}
    \item the segment contains a burst candidate, or
    \item the segment does not contain a burst candidate.
\end{enumerate}

Of course, a given data set may include receiver noise, RFI, bright individual pulses from pulsars, FRBs and other unexpected signatures.  RFI usually takes the form of wide-band, impulsive signals, or narrow-band, persistent signals, but can also mimic astronomical signals \citep{petroff15,men19}.

The classifier requires a training procedure.  Unfortunately, the number of known FRBs in the Parkes data sets is relatively low and single pulses from known pulsars all have relatively small DM values. This implies that we cannot simply train the algorithm on actual signals in the archival data. Instead, we inject simulated burst events into  1000 randomly chosen data files from the Parkes data archive. We simulated the bursts assuming the frequency-squared dispersion law. The FRB event is therefore parametrized by a time (corresponding to the arrival time of burst at the highest observing frequency), the DM, a width (the FRB is assumed to have a Gaussian profile)\footnote{ In the future we will re-train our model using a more physical parameterisation of the FRBs, including dispersion smearing, scattering, structures within the burst profile and the observed frequency dependence to the burst intensity. We note that our current bandwidth is relatively small (256\,MHz) and the channel bandwidths are relatively large (3\,MHz).  Scattering is usually small for FRB events and so the predominant effect is dispersion smearing (see, e.g., \citealt{2019ARA&A..57..417C}). However, dispersion smearing can be mitigated in searches for repeating events from known FRB sources as they can be carried out using coherently de-dispersed data streams.} and a brightness.  As we are injecting simulated FRBs into 1-bit data we need to ensure that we have a means to simulate different FRB brightnesses.  To do this, we define the fraction of samples within the FRB envelope that will become 1 (representing a signal above the mean level) and how many will remain 0 (a signal below the mean level). We note that no value that is already 1 will become a 0 in this process to ensure that any existing signal, such as RFI, is not affected by the simulation process.

For our training data set we simulated a wide-range of possible FRB parameters. The start time of the FRB was randomly chosen anywhere within the observation span, the DM values ranged between 20 and 5000 \,pc\,cm$^{-3}$, the saturation level measured by the percentage of pixels within the signal range turned bright by the FRB ranged from 75\% to 100\%, and the width of the FRB was chosen between 3 and 50 time samples.

\begin{table*}
	\centering
	\caption{The data files processed to develop and demonstrate our algorithm. We note that the testing data files are independent from the training data files.}
	\label{tab:collection_of_data}
	\begin{tabular}{lllrc p{2.4cm} ll}
		\hline
		Project	&	\multicolumn{1}{p{1.4cm}}{\centering Observing semester}	&	\multicolumn{1}{p{1.5cm}}{\centering Sampling time (s)}	&	\multicolumn{1}{p{0.7cm}}{\centering File count}	&	\multicolumn{1}{p{2.3cm}}{\centering Integration time per file (s)}	&	\multicolumn{1}{p{2.4cm}}{\centering Known signals}	&	\multicolumn{1}{p{2.9cm}}{\centering Reference}      &   \multicolumn{1}{p{1.4cm}}{\centering Usage}\\
		\hline
		P269	&	Jan 2001	&	0.001		&	2251	&	8400	&	Lorimer burst\protect{\newline} (FRB010724), and FRB010312	&	\cite{csiro:P269-2001JANT}  & training and testing\\
		P269	&	Oct 2000	&	0.001		&	951	&	2818	&	-						&	\cite{csiro:P269-2000OCTT}   & training\\
		P268	&	Aug 1997	&	0.00025		&	4615	&	2100	&	single pulses from known pulsars						&	\cite{csiro:P268-1997AUGT}  & training and testing\\
		P268	&	May 2001	&	0.00025		&	1820	&	2100	&	FRB010621					&	\cite{csiro:P268-2001MAYT}  &  testing\\
		\hline
	\end{tabular}
\end{table*}

We randomly selected 1000 files from data collections 1 to 3 in Table~\ref{tab:collection_of_data}, from which we extracted 57,000 data segments. We injected the simulated FRBs into 24,500 data segments. These resulting files therefore contained the real noise signals and our simulated FRBs. We have ensured that these specific data segments do not contain known FRBs or pulsars, but, of course, they may contain currently unknown, but real FRB events.  We also separately generated 7500 data segments in which we simulated a pure white-noise background and injected FRBs. These training data sets were used to encourage the model to learn the correct FRB patterns. These two sets of data segments form the positive training dataset while the remaining 32,500 data segments form the negative training dataset. 

\subsection{Deep Neural Networks Architecture}
\label{sec:DNN}

A detailed introduction of deep learning and related terminology can be found in~\cite{goodfellow2016deep}.  In the following, we make use of the following terms and concepts:
\begin{itemize}
    \item An image-based deep neural network (DNN) classifier, $\mathcal{F}(x;\theta)$, is a function that maps input image segments into a category, $ \hat{y} \in \{1, -1\}$, which indicates that the segment does, or does not, contain a signal of interest.
    \item $\mathcal{F}$ is a composite function, $\mathcal{F}(x; \theta) = f^l(f^{l-1}(...(f^2(f^1(x;\theta_1);\theta_2))...,\theta_{l-1});\theta_l)$, which contains multiple internal functions, $f^l$. $f^l$ is known as the $l$-th ``hidden layer'' of the network.
    \item Image segments are often enhanced prior to them being passed into a DNN classifier.  Methods such as applying a Gaussian filter are used to smooth the input images.
    \item It is important to determine how well a set of parameters models the given data. This is measured using a ``loss function'', $\mathcal{L}(\mathcal{F}(x; \theta), y)$,  which measures the performance of the function $\mathcal{F}(x; \theta)$, in which $y$ is the true label of $x \in X$.
    \item The DNN procedure obtains optimal values of the parameters $\theta$. An iterative method is used (we use the stochastic gradient descent algorithm) to minimise the loss function. This relies on a step size, known as the ``learning rate''.
    \item The composite function, $\mathcal{F}(x;\theta)$, contains various hidden layers (described above) including convolutional layers, max pooling layers and fully connected layers, etc. A convolutional layer can be thought of as a smoothing operation that applies on the input using a matrix often referred as ``kernel'' or ``filter''. The properties of the matrices are defined for specific features in the input images. Pooling layers reduce the dimensions of the data and hence simplify the computational complexity. Fully connected layers are directly connected to the inputs of the next layer. 
    \item Finally, the algorithm needs to convert the numerical values of the last layer to probabilities on the various possible classifications. A softmax function is often used for this.  
\end{itemize}

Our specific DNN is trained to identify single pulse events. $\mathcal{F}(x;\theta)$, is a function that maps our input image segment, $\{0,1\}^{96 \times 512}$, into a category $ \hat{y} \in \{1, -1\}$, which indicates that the segment does or does not contain an astronomical burst event respectively. We use the stochastic gradient descent algorithm  to optimize $\theta$ and fit the training data with a learning rate of $0.02$.  

To enhance the visual patterns within an image, each data segment is pre-processed using a Gaussian filter, which smooths the input data, before being fed into the network for training and prediction.   We have found that applying this pre-processing step improves the model training speed.

The first hidden layer in our $\mathcal{F}$  contains two parallel convolutional blocks. The first has a  $1 \times 1$ kernel with $8$ filters and the other uses a $9 \times 9$  kernel with $32$ filters. We use ReLU as the activation function for both blocks. The $1 \times 1$ kernel is introduced to add more non-linearity to the model in order to capture patterns of various forms of RFI in image segments. The $9\times9$ kernel is mainly introduced to capture the continuous patterns of the astronomical events and other non-astronomical signals. We apply a maximum pooling layer to the output of each convolutional block with a 2x2 patch size.
The output of the two maximum pooling layers are concatenated and fed into the second convolutional layer with a kernel size of $9 \times 9$ and 128 filters. We then apply another maximum pooling layer to the output of the second convolutional layer. The network stacks two more convolutional layers with a kernel size of $9 \times 9$ (with filter numbers of 256 and 512 respectively) and  maximum pooling layers before passing the output to a fully connected layer with 512 neurons. With a large number of parameters, DNN often overfits a training dataset (in particular with relatively few input examples).  We added a dropout layer to improve the generalization of  $\mathcal{F}$.  An additional fully connected layer with eight neurons is stacked before a softmax function is applied to obtain the probability distribution among the event and non-event categories. In order to improve the generalization of the model for different input data collections, we use L2 regularization. This regularizer makes the model avoid learning trivial features that only present in the training data.   The DNN classifier is implemented using Tensorflow.

We trained our neural network using the real and simulated data sets that were described in the previous section.  We then applied the trained model to data sets containing known events (such as the ``Lorimer burst'' and known pulsars) for demonstration and testing purposes.

\section{Saliency map analysis}

Saliency maps rank the pixels in the input image based on their influence on a probability score in a prediction~\citep{simonyan2013deep}. For deep neural networks, the influence can be calculated through the derivative of the score with respect to the input at the given pixel. To capture the variation of brightness of smoothed pixels in a pulse, we use ``integrated gradients'' \citep{Sundararajan:2017:AAD:3305890.3306024} to  distinguish the astronomical burst signature from background noise. We consider an input image, $x$, that
is formed by taking $n$ steps to add a value in each pixel from a black image $x'$ (each pixel has a value of 0). The integrated gradient of pixel $x_i$, denoted by $IG_i(x)$ is defined as below:
\begin{equation}
  IG_i(x)=(x_i - x'_i) \int_{\alpha=0}^1 \frac{\partial{\mathcal{F}(x'+\alpha (x - x')}}{\partial{x_i}}d\alpha
\label{equ:saliency}
\end{equation} 
in which $\mathcal{F}(x;\theta)$ is the DNN classifier and $\alpha$ denotes the step taken on the path of changing from $x'$ to $x$. The integrated gradients are able to determine how different pixels in the input image contribute to a prediction. 
 
\begin{figure}
    \centering
    \includegraphics[width=\columnwidth,viewport=10 0 705 680, clip]{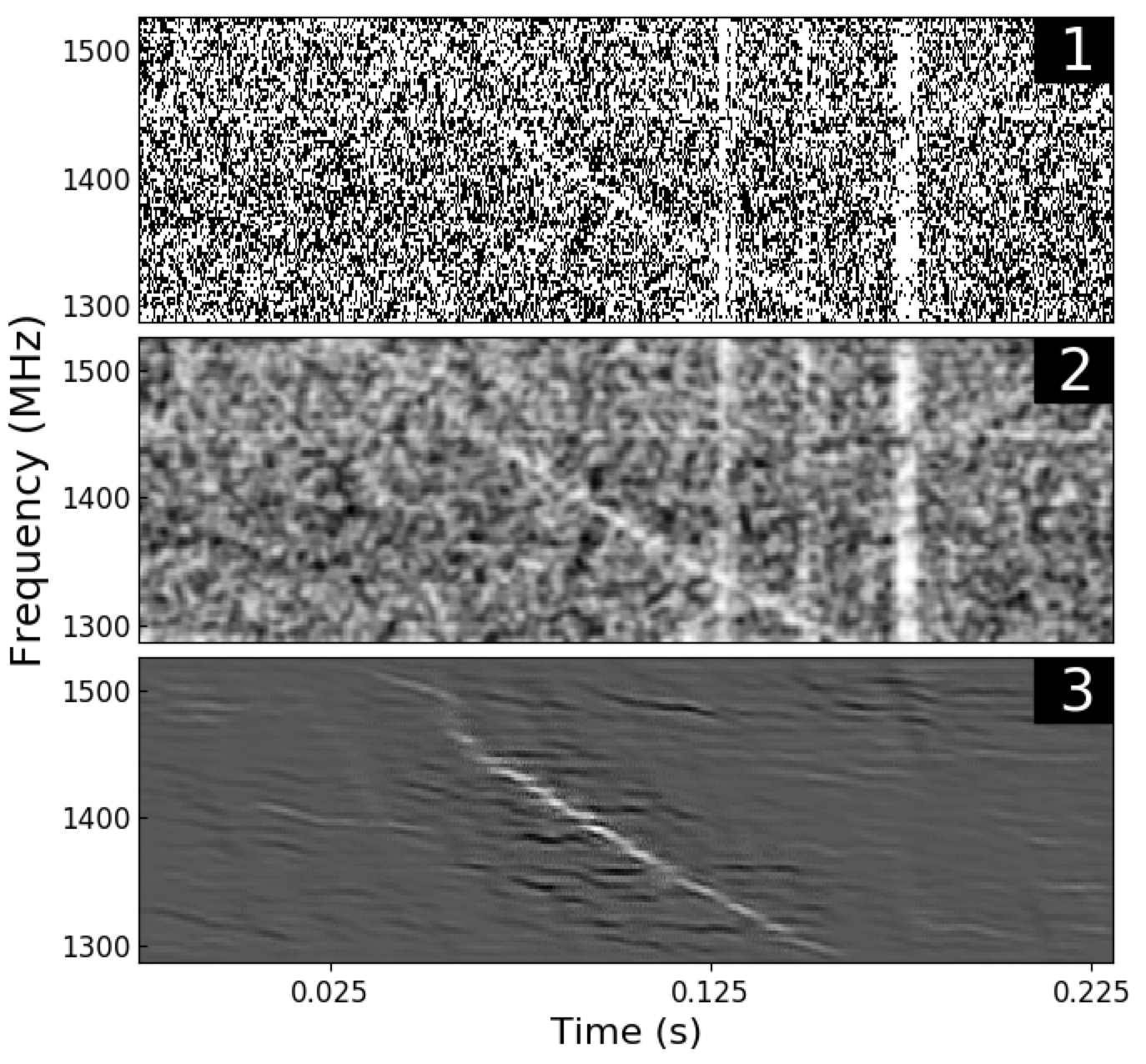}
    \caption{Demonstration of the use of saliency map to identify an individual pulse from the Vela pulsar (PSR~J0835$-$4510).  The upper panel shows the raw frequency-time image.  The curved feature is a single pulse from the pulsar. The vertical stripes are radio  interference. The central panel shows a smoothed version of the raw data. The saliency map is shown in the bottom panel.}
    \label{fig:saliency1} 
\end{figure}

To demonstrate this process, we provide an example in Figure~\ref{fig:saliency1}. For this observation the telescope was pointing towards the Vela pulsar, PSR~J0835$-$4510, and individual pulses from the pulsar are easily detectable (note that the Figure only shows a single pulse).  The top panel (labeled 1) is a segment of raw frequency-time data and clearly shows the pulse as well as three wide-band, impulsive RFI events.  We show the image after smoothing with a Gaussian filter in the middle panel (labeled 2).  Our machine learning classifier identified this region as containing an astrophysical event. However, we wish to ensure that it has correctly identified the pulse and not the RFI. The corresponding saliency map is shown in the bottom panel (labeled 3).  The brighter a pixel is in this panel, the more important it was when predicting that the data segment contains an FRB event. The classifier has correctly identified the single pulse as a feature for its positive classification, whilst ``ignoring'' the RFI features. 

\begin{figure}
    \centering
    \includegraphics[width=\columnwidth,viewport=10 0 705 680, clip]{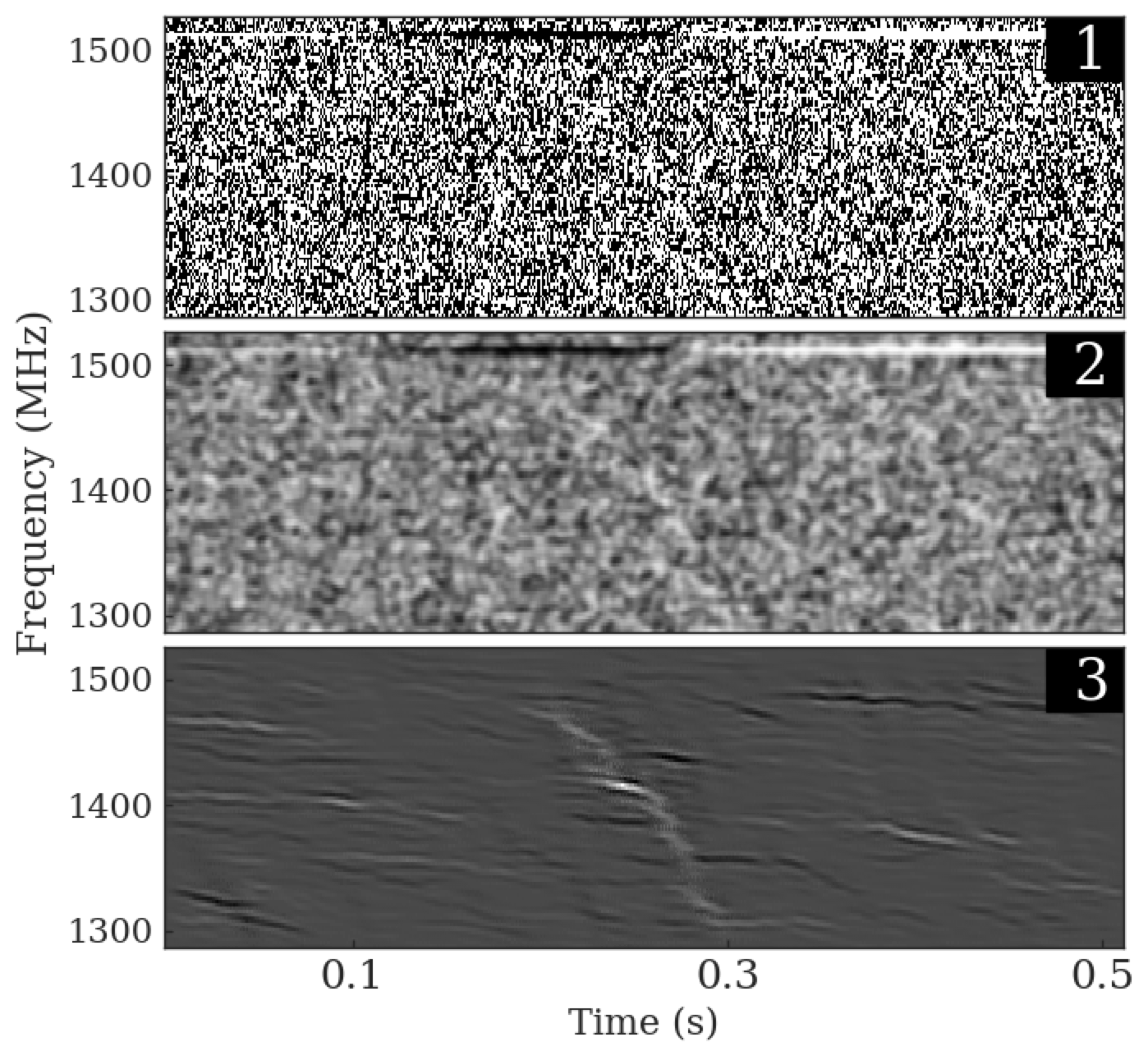}
    \caption{Demonstration of the use of saliency map to determine why a plausible event was not identified as an astrophysical source. The upper panel shows the burst event in the raw data. The central panel shows a smoothed version of the raw data. The saliency map is shown in the lower panel.}
    \label{fig:saliency2}
\end{figure}

Saliency maps can also be used for feature enhancement~\citep{simonyan2013deep} and to understand why a particular image \emph{was not} classified as an astrophysical burst event. For instance the image in Figure~\ref{fig:saliency2} was characterised as not containing any astrophysical burst event by the classifier.   Panel \#1 clearly shows narrow-band RFI around 1500\,MHz as well as a weak signal occurring between time $\sim$0.2 and 0.3\,sec.  This feature is not significantly enhanced in the raw data simply by smoothing the image (panel \#2). As we have defined the saliency map, bright pixels correspond to regions in the image that support the hypothesis than an FRB event is present.  The saliency map shows us that there is an FRB-like event present, but there is not sufficient evidence for the classifer to determine that this event is real.   This highlights the possibility of using saliency analysis to enhance features in images that have an FRB-like form, but are, in some way, different from the training data set (i.e., even though the algorithm was trained on ideal FRB events, the saliency maps can highlight similar, but not identical features).  

\begin{figure}
    \centering
    \includegraphics[width=\columnwidth,viewport=10 0 705 680, clip]{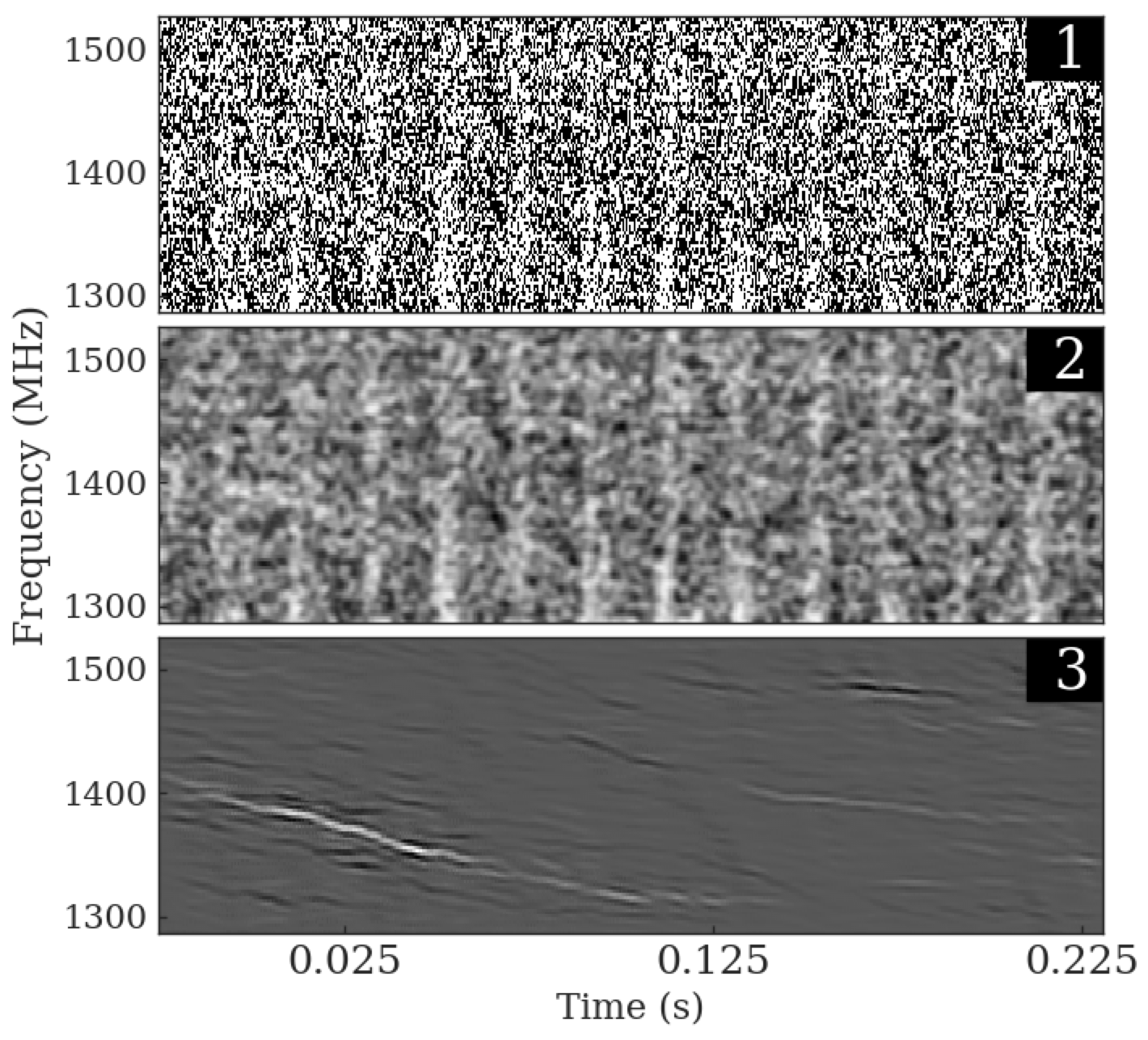}
    \caption{A single pulse event from PSR~J1536$-$5433 that is not centred in the image and affected by periodic radio frequency interference.}
    \label{fig:saliency3}
\end{figure}

In our first and second examples the events occured in the centre of the saliency map. This is coincidental and to demonstrate the success in saliency map analysis when the event is not directly in the centre of the image we show, in Figure~\ref{fig:saliency3}, an example in which the procedure clearly identifies and enhances a single pulse event from PSR~J1536$-$5433.  We emphasise that no RFI rejection was carried out and yet the impulsive, broadband RFI clearly present in the raw data is not present in the saliency map.

In order to explore the use of saliency maps further, we have analysed a sub-set of our real and simulated data files using both a traditional PRESTO single pulse search pipeline \citep{2001PhDT.......123R} and our ML algorithm. The PRESTO pipeline has been described in detail by \cite{2018MNRAS.479.1836Z} and \cite{2019MNRAS.484L.147Z}. In brief, it uses {\tt rfifind} to mask strong narrow-band and short-duration broad-band RFI. The {\tt -noclip} option is turned on to avoid deleting potential bursts. {\tt DDplan} is then used to determine the DMs to be trialled in the dedispersion phase (set here to have 440 trials between 0 and 5000 \,pc\,cm$^{-3}$). Dedispersion is carried out using {\tt prepdata} and RFI removed using the masks produced by {\tt rfifind}.  

A direct comparison of candidate lists is non-trivial and an in-depth comparison between methods and candidates will be presented elsewhere. One challenge in comparing candidate lists is that the PRESTO pipeline often produces multiple candidates for the same event (with slightly different DMs, event times and or widths), another challenge is that the PRESTO pipeline is extremely versatile and can be ``tuned'' using different input parameters.  Our PRESTO-based pipeline grouped all the candidates that occurred close together in time (within a 10\,ms time window). If the candidate with the highest S/N in a group has a S/N greater than seven then it was manually inspected.  We found that there was an exact match between candidate lists for candidates with S/N $>$ 12.  These particular candidates were single pulses from known pulsars and the agreement with PRESTO shows that the ML algorithm is sufficient for the tests we describe here.  We note that our ML algorithm is significantly faster than the PRESTO search as our process contains no de-dispersion, nor RFI mitigation stages.

The RFI-free and feature-enhanced saliency images can be used to enable a direct fit for the DM of the event and to get an estimate of the significance of the event in a way that is not affected by RFI (and without any de-dispersion steps).   To show this we have determined the S/N for four examples shown in this paper as determined using the PRESTO pipeline~\citep{2018MNRAS.479.1836Z,2019MNRAS.484L.147Z} and compared those results with those obtained from a fit to the event in the saliency map. The raw data containing PSR~J0835$-$4510 and PSR~J1536$-$5433 contain impulsive, broadband RFI whereas the Lorimer burst, PSR~J1057$-$5226 and PSR~J1744$-$3130 data sets have little detectable RFI.   The S/N values are listed in Table~\ref{tb:comparison}, where S/N$_1$ is the signal to noise determined from the PRESTO pipeline and S/N$_2$ from our pipeline. Comparison is non-trivial as the noise is well defined for the traditional analysis, but the noise in a saliency map is non-Gaussian.  However, we can clearly see that the S/N$_1$ values range from 4 (for an data set affected by RFI) to 35 (for a bright event in a clean data set), whereas all the S/N$_2$  values are similar.

\begin{table}

\caption{Comparison between determining S/N\tnote{a} of using the raw data file and the saliency map. }
\begin{threeparttable}
\begin{tabular}{p{2.2cm}llllll}
\hline

Event & S/N$_1$ & nbin$_1$ & S/N$_2$ & nbin$_2$ & DM \\
      &         &          &         &          &(cm$^{-3}$pc) \\
\hline
Lorimer burst & 35.18 & 26 & 55.06 & 14 & 375 \\
PSR~J0835$-$4510 & 10.54 & 10 & 58.69 & 14 & 67.97  \\
PSR~J1536$-$5433 & 4.27 & 12 & 67.16 & 18 & 147.5  \\
PSR~J1057$-$5226 & 13.08 & 6 & 49.46 & 14 & 29.69  \\
PSR~J1744$-$3130 & 6.86 & 18 & 37.15 & 20 & 192.90  \\
\hline
\hline
\end{tabular}
    \begin{tablenotes}
          \footnotesize
           \item[a]We use the definition of ${\rm \sigma}$ = $\sum {(\rm signal - noise})$/${\rm RMS}$/$\sqrt{\rm nbin}$  for our S/N value, where ${\rm nbin}$ is the best filter width in units of samples. 
    \end{tablenotes}
\end{threeparttable}
\label{tb:comparison}
\end{table}

\begin{figure}
	\subfigure[]{
	\begin{minipage}{\columnwidth}
		\centering
		\includegraphics[width=0.95\columnwidth,viewport=0 0 670 540, clip]{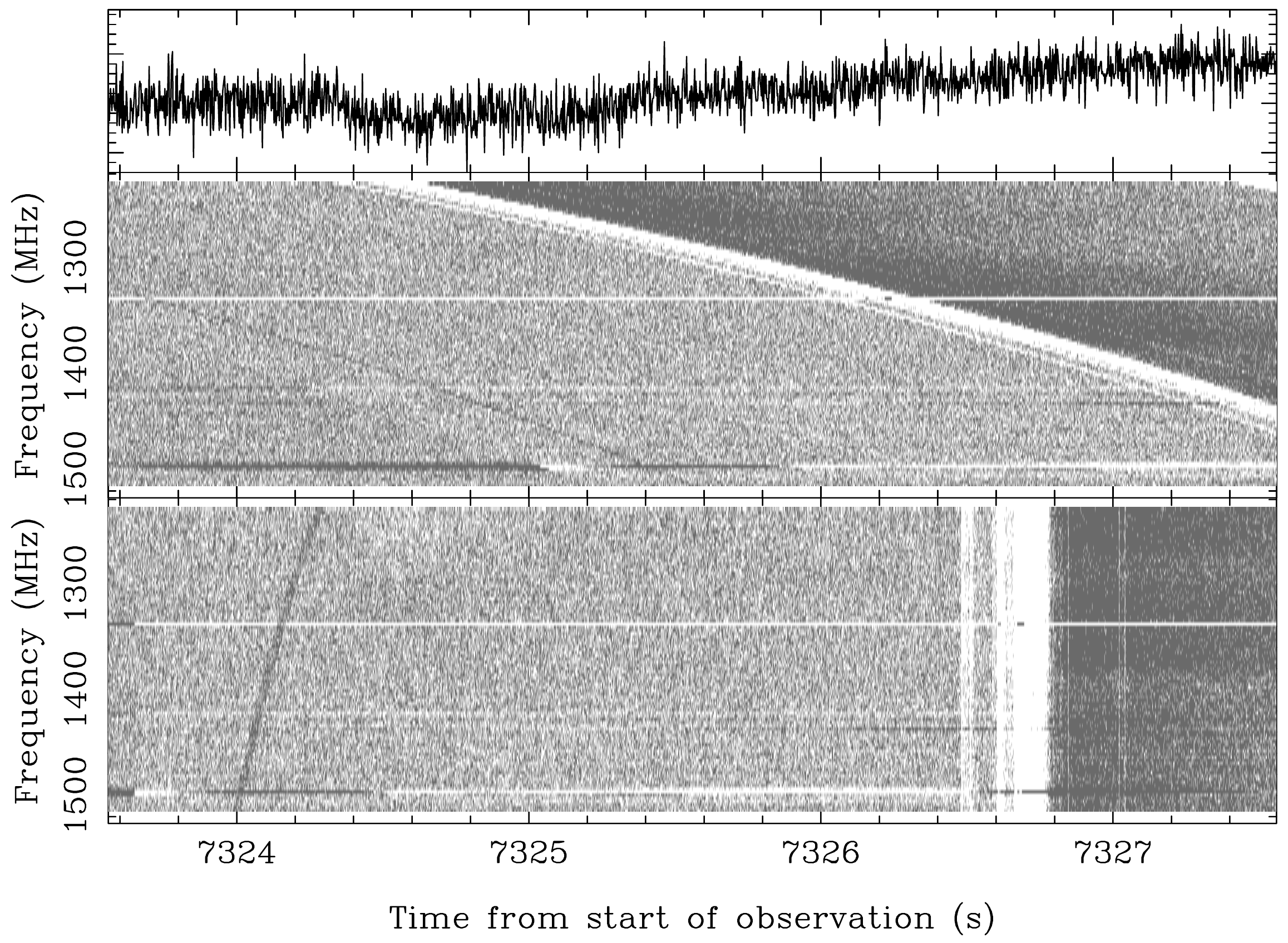}
\begin{picture}(0, 0)
\linethickness{2pt}
\put(-210,26){\framebox(35,52){}}
\end{picture}
		\label{fig:fn-presto1}
    \end{minipage}}

	\subfigure[]{
	\begin{minipage}{\columnwidth}
		\centering
		\includegraphics[width=0.95\columnwidth,viewport=0 0 670 490, clip]{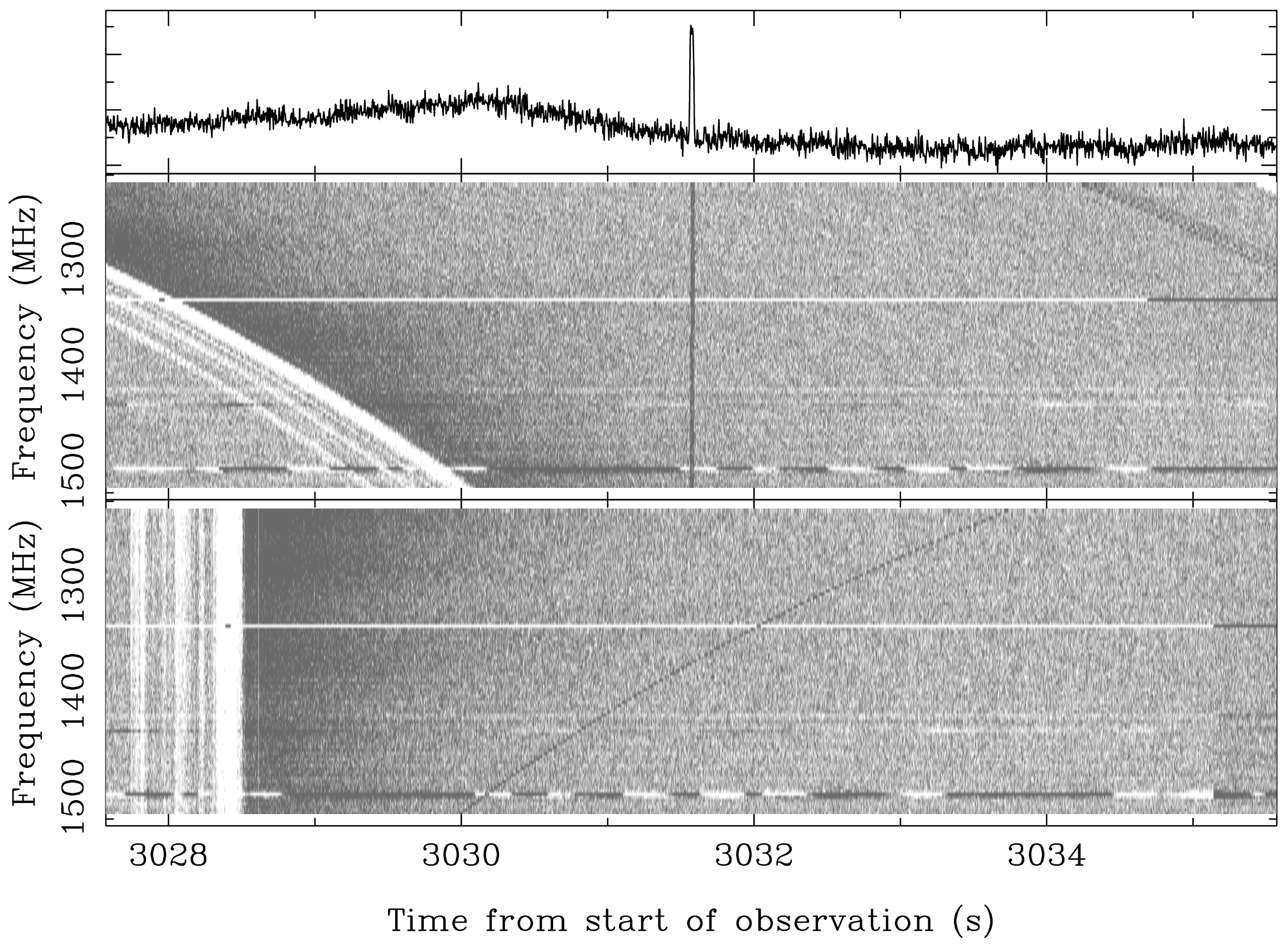}
		\label{fig:fn-presto3}
	\end{minipage}}

    \caption{False negative examples from the PRESTO single pulse search pipeline on data sets in which fake FRB signals have been injected. The sub-panels are described in the text.  In (a) the candidate detected by the PRESTO pipeline has a DM of 3962 \,pc\,cm$^{-3}$ whereas the actual signal has a DM  of 300 \,pc\,cm$^{-3}$. 
    In (b) the S/N of a signal is distorted by RFI, which leads to a low S/N ratio for the FRB event.} 
	\label{fig:FNpresto}
\end{figure}

The ML algorithm produced fewer candidates than PRESTO for events with S/N$_1$ values $<$12.  A future in-depth analysis will determine whether this is because the ML algorithm is missing true candidates, or whether PRESTO is presenting false positives.  We also found that the PRESTO-based pipeline did not identify a few of our simulated FRB events that were injected into actual data sets (i.e., these are false negative examples). These events were successfully detected using the ML method.   In Figure~\ref{fig:FNpresto}, we show two such false negative examples. We note that the panels in these images differ from the saliency map demonstrations. Here the top panel shows the dedispersed time series (using the DM of the PRESTO candidate), the central panel shows the dedispersed data as a function of time and frequency and the bottom panels shows the raw data. In Figure~\ref{fig:fn-presto1}, the candidate produced by the PRESTO pipeline has a DM of 3962\,pc\,cm$^{-3}$, whereas the simulated signal (highlighted with a black rectangle) has a DM of 300 \,pc\,cm$^{-3}$. Our choice of parameters when running the PRESTO pipeline failed to detect (and remove) the RFI that shadows the real signal (the  S/N of the RFI is 242.19). Our machine learning algorithm correctly identified this simulated FRB event with a S/N of 25.9. In Figure~\ref{fig:fn-presto3}, the measurement of the S/N of the simulated signal is affected by RFI. This candidate was found by PRESTO with a low S/N value and was filtered out by our S/N cutoff. The machine learning algorithm correctly detects this signal with a S/N of 29.0.  

\begin{figure}
    \centering
	\includegraphics[width=\columnwidth,viewport=10 0 705 680, clip]{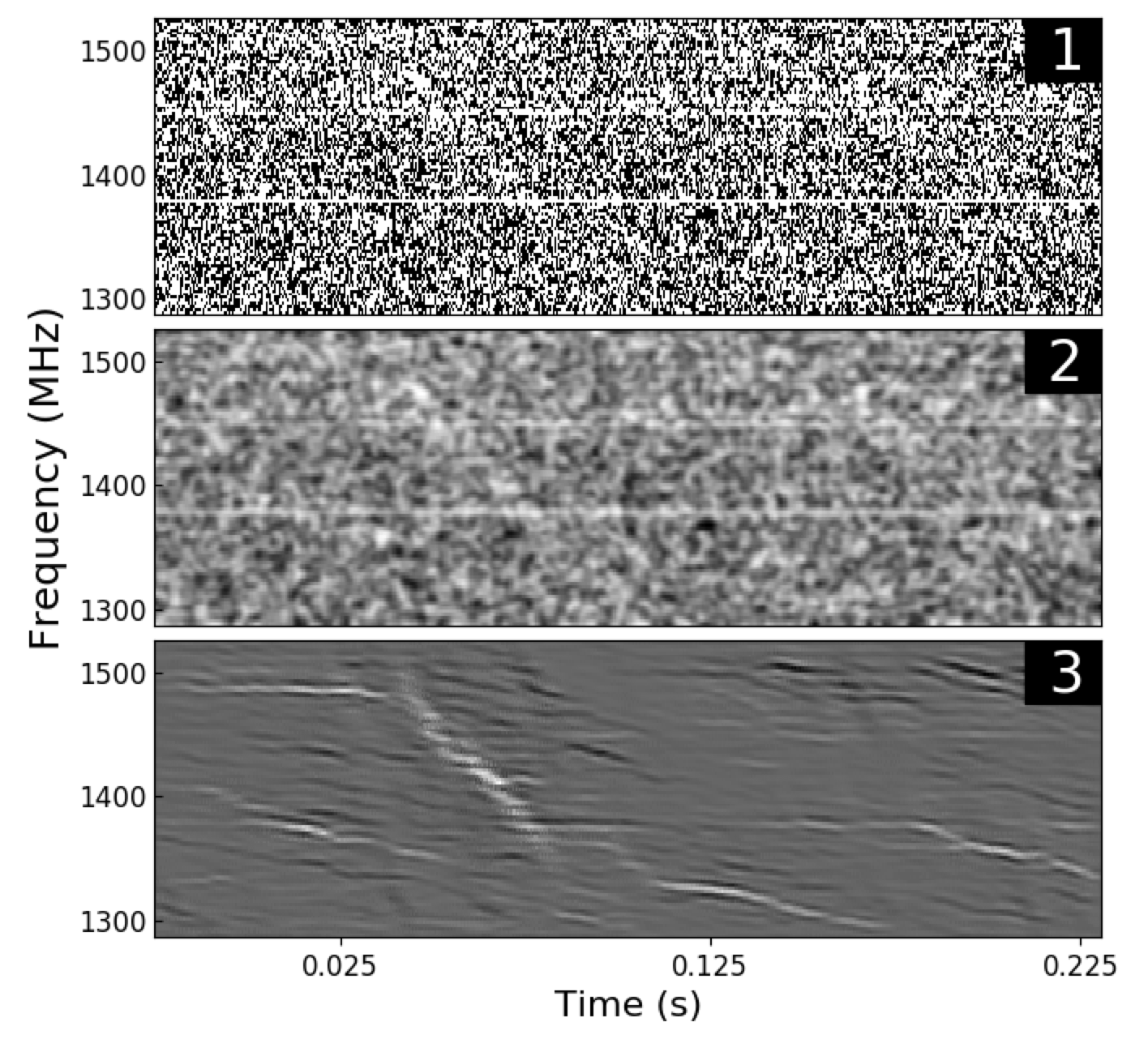}
    \caption{An unexplained transient signal with a dispersion measure of 41\,pc\,cm$^{-3}$. The raw data are shown in the upper two panels. The saliency map image is presented in the lower panel.}                                      
    \label{fig:candidate}
\end{figure}

 During the testing of our classifier we found a new single pulse from an unknown source from the observation file PM0143\_012D1.sf in the data collection P268--2001MAY. In Figure~\ref{fig:candidate}, we present the raw data and the corresponding saliency map for this potential discovery, which has a right ascension and declination of 19:14:43 and $+$02:26:13 respectively. Panel 1 contains the raw time-frequency data, which is smoothed in panel 2. The saliency map is given in panel 3. The DM corresponding to this candidate is 41 \,pc\,cm$^{-3}$ which is significantly smaller than the Galactic contribution in the source direction. The traditional pulses search pipeline (such as PRESTO;~\citealt{2001PhDT.......123R}) identified this event with a S/N of $\sim 7$, but the S/N in the saliency map is 23.3 and we see no other comparable unknown event in our processing.  If real (and further observations are planned of this sky region), then the source will be from a currently-unknown pulsar or a rotating radio transient (RRAT; \citealt{2006Natur.439..817M}). 

\section{Conclusions}

The primary goal of this paper is to highlight that saliency map analysis can be used to provide confidence that a machine learning algorithm has identified a real astronomical event in a given data set.  Producing saliency maps is not computationally intensive. Our initial implementation takes only $\sim$1\,s to form a saliency map for a given candidate (and this can be further optimised).  For this demonstration we have made use of archival data that have been 1-bit digitised.  The results presented here are general and are directly applicable to higher-bit data streams. 

We note that saliency maps are not unique to high time resolution data sets.  Source finding is being carried out in  interferometric images to search for unusual source lobes and jets \citep{2019PASP..131j8004N}. Saliency map analysis can be used to enhance such features in such complex images.

In summary, we have developed a new machine learning procedure for identifying astrophysical burst events in time-domain data streams. A detailed analysis of our algorithm and how it compares with more traditional search method will be published elsewhere.  Here, we have explored the use of saliency maps to identify the signatures within the data stream that contain the burst event.  We have shown that the saliency maps are robust in the presence of RFI and provide a method to enhance burst-like signatures in a given data stream.  

With the advent of new telescopes and improved instrumentation, the ability to detect burst events using traditional methods will become harder. The enormous data volumes from, for example, the Five-hundred-meter Aperture Spherical radio Telescope (FAST, \citealt{2012ApJ...759..127M}; \citealt{2018IMMag..19..112L}) and the Square Kilometre Array (SKA) will
require that computationally-efficient algorithms
be developed and it will not be possible to view every candidate by eye. Machine learning algorithms, such as the one described here, clearly have a role to play in extracting the astrophysical information from such large data sets.

\bibliographystyle{aa}
\bibliography{frb}

\begin{acknowledgements}
This work is supported by the National Natural Science Foundation of China (Grant No. U1731238, 11725313, 11690024, 11743002, 11873067, U1731218, 11565010, 11603046, U1531246, 11703047, 11590783) and the National Key R\&D Program of China No. 2017YFA0402600 and the CAS ``Light of West China'' Program. The Parkes radio telescope is part of the Australia Telescope National Facility which is funded by the Australian Government for operation as a National Facility managed by CSIRO. This paper includes archived data obtained through the CSIRO Data Access Portal (http://data.csiro.au). This work was supported by a China Scholarship Council (CSC) Joint PhD Training Program grant and the National Natural Science Foundation of China. This project was supported by resources and expertise provided by CSIRO IMT Scientific Computing.
\end{acknowledgements}

\end{document}